\pdfoutput=1
\documentclass[10pt,aps,prl,twocolumn,groupedaddress,longbibliography]{revtex4-1}
\usepackage{amsmath,amssymb,epsfig,MnSymbol}
\usepackage[colorlinks=true,citecolor=blue]{hyperref}

\begin{document}


\title{Universal Coherence-Induced Power Losses of Quantum Heat 
Engines\\ in Linear Response}
\author{Kay Brandner\textsuperscript{1}}
\author{Michael Bauer\textsuperscript{2}}
\author{Udo Seifert\textsuperscript{2}}
\affiliation{\textsuperscript{{{\rm 1}}}Department of Applied Physics,
Aalto University, 00076 Aalto, Finland\\
\textsuperscript{{{\rm 2}}}II. Institut 
f\"ur Theoretische Physik, Universit\"at Stuttgart, 70550 Stuttgart, 
Germany}




\begin{abstract}
We introduce a universal scheme to divide the power output of a 
periodically driven quantum heat engine into a classical contribution
and one stemming solely from quantum coherence.  
Specializing to Lindblad-dynamics and small driving amplitudes, we 
derive general upper bounds on both, the coherent and the total power.
These constraints imply that, in the linear-response regime, coherence
inevitably leads to power losses. 
To illustrate our general analysis, we explicitly work out the
experimentally relevant example of a single-qubit engine.
\end{abstract}


\maketitle

\vbadness=10000
\hbadness=10000

\newcommand{\matt}[4]{\left(\!\begin{array}{ll}
#1 & #2\\ #3 & #4 \end{array}\!\right)}

\newcommand{\Sp}[1]{\left\langle #1 \right\rangle}
\newcommand{\Spb}[1]{\bigl\langle #1 \bigr\rangle}
\newcommand{\Spn}[1]{\langle #1 \rangle}
\newcommand{\SpB}[1]{\biggl\langle #1 \biggr\rangle}
\newcommand{\abs}[1]{\left| #1\right|}
\newcommand{\tr}[1]{{{{\rm tr}}}\!\left\{ #1 \right\}}
\newcommand{\ket}[1]{\left|#1\right\rangle}
\newcommand{\bra}[1]{\left\langle #1\right|}
\newcommand{\shortsum}[1]{\sum\mathop{}_{\mkern-5mu #1}}
\newcommand{\braket}[2]{\left\langle\left.#1\right| #2\right\rangle}

\newcommand{\x}{{{{\rm x}}}}
\renewcommand{\r}{\varrho}
\renewcommand{\j}{\mathbf{j}}
\renewcommand{\k}{\mathbf{k}}
\newcommand{\ddg}{\ddagger}

\newcommand{\kb}{k_{{{\rm B}}}}
\newcommand{\T}{\mathcal{T}}
\renewcommand{\tint}{\int_0^\T\!\!\! dt \;}
\newcommand{\tauint}{\int_0^\infty\!\!\! d\tau \;}
\newcommand{\taupint}{\int_0^\infty\!\!\! d\tau'\;}
\newcommand{\ttauint}{\int_0^\T \!\!\! dt \!
                      \int_0^\infty\!\!\! d\tau\;}

\newcommand{\K}{\mathsf{K}}
\newcommand{\s}{\sigma}
\newcommand{\vt}{\vartheta}
\newcommand{\D}{\mathsf{D}}
\renewcommand{\H}{\mathsf{H}}
\renewcommand{\L}{\mathsf{L}}
\newcommand{\Gc}{\delta G^{{{\rm c}}}}
\newcommand{\Gs}{\delta G^{{{\rm d}}}}
\newcommand{\Gq}{\delta G^q}
\newcommand{\Gw}{\delta G^w}

\newtheorem{lem}{Lemma}
\newtheorem{cor}{Corollary}

\begin{figure}[!ht]
\flushleft
\includegraphics[scale=1.31]{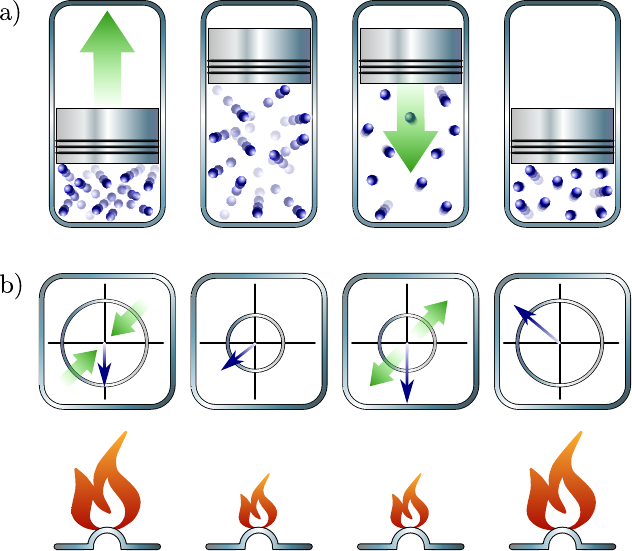}
\caption{Classical and quantum engines. 
a) Macroscopic Stirling cycle. 
In the first stroke, mechanical power is extracted by expanding the
hot working fluid.
Decreasing the temperature at constant volume in the second stroke
leads to a reduction of pressure before the gas is compressed again in
the third stroke. 
The cycle is completed by isochorically returning to the initial 
temperature. 
b) Quantum Stirling cycle. 
The working fluid consists of a two-level system, whose Bloch vector
is shown in the four diagrams corresponding to the beginning of each
stroke. 
Coordinates are chosen such that the instantaneous energy eigenstates
lie on the vertical axis. 
The radius of the circle is proportional to the level splitting.
Two distinct control operations are used to realize the work strokes:
the level splitting is changed externally and superpositions between
the two levels are created, i.e., the Bloch vector is rotated away
from the vertical axis. 
During the thermalization strokes, coherence is irreversibly
destroyed and the level population adapts to the temperature of the
environment. 
\label{FigEnConv}}
\end{figure}

Heat engines are devices that convert thermal energy into useful 
work. 
A Stirling motor, for example, uses the varying pressure of a 
periodically heated gas to produce mechanical motion, 
Fig.~\ref{FigEnConv}a.
Used by macroscopic engines for two centuries, this elementary 
operation principle has now been implemented on ever-smaller scales. 
Over the last decade, a series of experiments has shown that the 
working fluid of Stirling-type engines can be reduced to tiny 
objects such as a micrometer-seized silicon spring
\cite{Steeneken2010} or a single colloidal particle \cite{Blickle2011,
Martinez2015,Martinez2015a,Krishnamurthy2016}.
These efforts recently culminated in the realization of a single-atom
heat engine \cite{Roßnagel2015,Abah2012}.
Thus, the dimensions of the working fluid were  further decreased 
by four orders of magnitude within only a few years. 
In light of this remarkable development, the challenge of even smaller
engines operating on time and energy scales comparable to Planck's 
constant appears realistic for future experiments. 

Such quantum engines would have access to a non-classical mechanism of
energy conversion that relies on the creation of coherent 
superpositions between the energy levels of the working fluid
\cite{Uzdin2015}, Fig.~\ref{FigEnConv}b.
How does this additional freedom affect performance figures like power
and efficiency? 
Having triggered substantial research efforts in recent years, 
this question constitutes one of the central problems in the 
emerging field of quantum thermodynamics, see for example 
\cite{Quan2007,Funo2013,Rossnagel2014,Pekola2015,Zhou2015,Lostaglio2015,
Cwikli2015,Gardas2015b,Tasaki2016a,Jaramillo2016}.
However, the available results are so far inconclusive. 
In fact, current evidence suggests that, depending on the specific 
setup and benchmark parameters, coherence can, in principle, be both
conducive \cite{Scully2011,Abe2012,Correa2014,Horowitz2014,
Rossnagel2014,Brandner2015d,Mitchison2015a,Gelbwaser-Klimovsky2015b,
Uzdin2015,Uzdin2016b,Watanabe2016a} and detrimental \cite{Feldmann2003,
Feldmann2004,Brandner2016,Roulet2016,Karimi2016a} to the performance of
thermal devices.

In this article, we universally characterize the role of coherence for
the power output of cyclic heat engines in linear response.
Our analysis builds on the well-established theory of open quantum
systems \cite{Alicki1979b,Breuer2006} and a recently developed 
thermodynamic framework describing periodically driven systems
\cite{Brandner2015f,Brandner2016}, which has already proven very
useful in the classical realm \cite{Proesmans2015a,Bauer2016a,
Proesmans2016b,Proesmans2016c}.
For a quantum engine, we model the working fluid as an $N$-level 
system with bare Hamiltonian $H$, which is embedded in a large 
reservoir with base temperature $T$ \cite{Kosloff2013,Brandner2016}. 
For simplicity, we assume that $N$ is finite.
A heat source injects thermal energy into the system by periodically
heating its local environment. 
Hence, the working fluid effectively feels the time-dependent 
temperature
\begin{equation}
T_t\equiv T+ f^q_t,
\end{equation}
where $f^q_t\geq 0$. 
For work extraction, a periodic driving field $f^w_t$ is applied,
which couples linearly to the degree of freedom $G_w$ of the system. 
The Hamiltonian thus acquires the time-dependence
\begin{equation}\label{Hamiltonian}
H_t = H + f^w_t G^w.
\end{equation}
For uniqueness, we assume that the field $f^w_t$ is dimensionless and 
that its average over one period $\T$ vanishes. 

This engine delivers the mean power output
\begin{equation}\label{PowTot}
P = -\frac{1}{\T}\tint\tr{\dot{H}_t \r_t},
\end{equation}
where $\r_t$ denotes the periodic state of the system 
\cite{Alicki1979b}.
Using the spectral decomposition 
\begin{equation}\label{SpectDecomp}
H_t \equiv \sum_n E^n_t \ket{n_t}\bra{n_t}
\end{equation}
of the time-dependent Hamiltonian, $P$ can be divided into two 
contributions corresponding to the different mechanisms of work 
extraction illustrated in Fig.~\ref{FigEnConv}.
First, the classical power 
\begin{equation}\label{PowClass}
P^{{{\rm d}}}\equiv -\frac{1}{\T}\tint \sum_n \dot{E}^n_t 
                       \bra{n_t}\r_t\ket{n_t}
\end{equation}
is generated by changing the energy levels of the working fluid, i.e.,
the diagonal elements of its Hamiltonian with respect to the 
unperturbed energy eigenstates. 
Second, the coherent power 
\begin{equation}\label{PowQuant}
P^{{{\rm c}}}\equiv P-P^{{{\rm d}}} =
\frac{1}{\T}\tint \sum_n \bra{\dot{n}_t}[H_t,\r_t]\ket{n_t}
\end{equation}
arises from creating superpositions between the instantaneous energy
eigenstates
\footnote{To make these definitions unique, we 
understand that the time-dependent energy eigenvalues are arranged in
increasing order, i.e., $E^0_t\leq E^1_t\leq \cdots\leq E^N_t$ for
any $t\in\mathbb{R}$.
Furthermore, we note that the expressions \eqref{PowClass} and 
\eqref{PowQuant} are unaltered if the instantaneous energy 
eigenstates are multiplied with arbitrary time-dependent phase
factors.}.
Accordingly, $P^{{{\rm c}}}$ vanishes when $\r_t$ commutes with 
$H_t$ throughout one operation cycle.
This condition is met, for example, in the adiabatic limit, where 
the state of the system follows the instantaneous Boltzmann
distribution. 

For deriving constraints on the coherent power $P^{{{\rm c}}}$, we 
have to specify the dissipative dynamics of the working fluid.
To this end, we invoke the standard condition of weak coupling 
between system and reservoir. 
In equilibrium, i.e., for $f^q_t=f^w_t=0$, the state $\r_t$ evolves
according to the Markovian master equation \cite{Breuer2006}
\begin{equation}\label{MasterEq}
\partial_t\r_t = -\frac{i}{\hbar}[H,\r_t] + \D\varrho_t,
\end{equation}
where the dissipator 
\begin{equation}\label{Dissipator}
\D X\equiv  \sum_\sigma \frac{\gamma_\sigma}{2}\left(
                  [V_\sigma X,V_\sigma^\dagger]
                 +[V_\sigma,X V_\sigma^\dagger]\right)
\end{equation}
accounts for the influence of the thermal environment 
\cite{Spohn1978a}.
Furthermore, $\hbar$ denotes Planck's constant and 
$\{\gamma_\sigma\}$ is a set of positive rates with corresponding 
Lindblad operators $\{V_\sigma\}$.
Due to microreversibility, these quantities are constrained by the
quantum detailed balance relation, which can be expressed compactly 
in terms of the formal identity \cite{Alicki1976,Spohn1978a}
\begin{equation}\label{DB}
\D e^{-\beta H} = e^{-\beta H} \D^\dagger.
\end{equation}
Here, $\beta\equiv 1/(\kb T)$, $\kb$ denotes Boltzmann's constant and
the adjoint dissipator is given by 
\begin{equation}\label{AdjDissipator}
\D^\dagger X\equiv\sum_\sigma \frac{\gamma_\sigma}{2}
\left(V_\sigma^\dagger[X,V_\sigma]+ 
     [V_\sigma^\dagger,X]V_\sigma\right). 
\end{equation}
 
Provided that the cycle period $\T$ is large compared to the 
relaxation time of the reservoir, finite driving can be included in
this framework by allowing the rates and Lindblad operators to be 
time-dependent and replacing $H$ and $T$ with $H_t$ and $T_t$
respectively in \eqref{MasterEq}-\eqref{AdjDissipator}
\cite{Alicki1979b}.
Solving the master equation \eqref{MasterEq} by treating $f^q_t$ and 
$f^w_t$ as first-order perturbations then yields the explicit 
expressions \footnote{To obtain \eqref{PowExpl} from \eqref{PowClass}
and \eqref{PowQuant}, standard linear-response theory is applied. 
This derivation exploits that, due to the detailed balance relation
\eqref{DB}, the space of all system operators commuting with the
unperturbed Hamiltonian $H$ is invariant under the action of the 
super operators $\D$ and $\D^\dagger$, for details see 
\cite{Brandner2016}.}
\begin{align}\label{PowExpl}
P^{{{\rm d}}}&\equiv -\frac{1}{\T}\ttauint\dot{f}^w_t\left(
                   \dot{C}^{{{\rm dd}}}_\tau f^w_{t-\tau} +
                   \dot{C}^{{{{\rm d}}} q}_\tau f^q_{t-\tau}
                   \right)\quad\text{and}
\nonumber\\
P^{{{\rm c}}}&\equiv -\frac{1}{\T}\ttauint\dot{f}^w_t
                   \dot{C}^{{{\rm cc}}}_\tau f^w_{t-\tau} 
\end{align}
for the classical and the coherent power, respectively
\footnote{The detailed-balance relation \eqref{DB} implies that the
set of Lindblad operators $\{V_\sigma\}$ is self-adjoint
\cite{Alicki1976,Kossakowski1977}. 
Additionally, we here assume that this set is irreducible such that 
$X=1$ is the only solution of $\D^\dagger X = 0$
\cite{Spohn1977}.  
Under this condition, the improper integrals showing up in 
\eqref{PowExpl} are well-defined \cite{Brandner2016}}, in the 
following notation. 
We abbreviate with $C^{ab}_t$ the Kubo correlation function
\cite{Kubo1998}
\begin{equation}\label{CorrFunDef}
C^{ab}_t\equiv
\bigl\llangle \hat{G}^a_t,\hat{G}^b_0\bigr\rrangle \equiv
\int_0^\beta \!\!\! d\lambda \;\left(
\Sp{\hat{G}^a_t e^{-\lambda H} \hat{G}^b_0 e^{\lambda H}}
-\Sp{\hat{G}^a_t}\Sp{\hat{G}^b_0}\right),
\end{equation}
where $t\geq 0$, $a,b={{\rm d}},{{\rm c}},q$. 
Hats indicate Heisenberg-picture operators satisfying the adjoint
master equation 
\begin{equation}
\partial_t\hat{X}_t = \frac{i}{\hbar}[H,\hat{X}_t]
                    + \D^\dagger\hat{X}_t
\end{equation}
with initial condition $\hat{X}_{0}=X$ \cite{Breuer2006}. 
The angular brackets in \eqref{CorrFunDef} denote the thermal average,
i.e., 
\begin{equation}
\Sp{X} \equiv \tr{X e^{-\beta H}}/\tr{e^{-\beta H}}.
\end{equation}
Finally, we have defined the operator $G^q\equiv -H/T$ and split 
the control variable $G^w$ into a diagonal, quasi-classical, and a
coherent part,
\begin{equation}
G^{{{\rm d}}}\equiv \sum_n \ket{n}\bra{n}G^w\ket{n}\bra{n}
\quad\text{and}\quad
G^{{{\rm c}}}\equiv G^w-G^{{{\rm d}}},
\end{equation}
where the vectors $\ket{n}$ correspond to the eigenstates of 
the unperturbed Hamiltonian $H$. 

As a first key-observation, we note that the expression 
\eqref{PowExpl} for $P^{{{\rm c}}}$ is independent of the temperature
profile $f^q_t$.
Thus, under linear-response conditions, it is impossible to convert
thermal energy provided by the heat source into positive power 
output via quantum coherence; 
rather coherent power can only be injected into the system through
mechanical driving.
This constraint is captured quantitatively by the bound
\begin{equation}\label{PowCBound}
P^{{{\rm c}}}\\
\leq -\frac{L^{{{\rm c}}}_1\Omega^2}{\Omega^2+L_2^{{{\rm c}}}/
L^{{{\rm c}}}_1} F^w\leq 0,
\end{equation}
which is saturated in the two limits $\Omega\rightarrow 0$  and 
$\Omega\rightarrow\infty$, for the proof see \cite{SM}.
Besides the cycle frequency $\Omega\equiv 2\pi/\T$, the bound 
\eqref{PowCBound} involves the mean square amplitude 
\begin{equation}\label{WorkProtSqAmpl}
F^w\equiv \frac{1}{\T}\tint (f^w_t)^2
\end{equation}
of the driving field, and the Green-Kubo type coefficients 
\begin{equation}\label{GreenKuboCoeffC}
L^{{{\rm c}}}_j\equiv\int_0^\infty \!\!\! dt\;
\bigl\llangle \hat{G}_t^{{{\rm c}}(j)},
\hat{G}_0^{{{\rm c}(j)}}\bigr\rrangle\geq 0,
\end{equation}
where the index $j$ in brackets means a time-derivative of respective 
order. 

The bound \eqref{PowCBound} can be understood intuitively by 
identifying  the parameter $L^{{{\rm c}}}_2/L^{{{\rm c}}}_1$ as an 
estimator for the decoherence strength of the reservoir, i.e.,
the square of the mean rate, at which its influence destroys coherent
superpositions between the energy levels of the working fluid.
In the incoherent limit $L^{{{\rm c}}}_2/L^{{{\rm c}}}_1\gg\Omega^2$,
the coherent power can approach zero due to frequent interactions with
the environment constantly forcing the system into a state that is
diagonal in the instantaneous energy eigenbasis. 
This behavior resembles the quantum Zeno effect with the role of the
observer played by the thermal reservoir \cite{Breuer2006}. 
If $L^{{{\rm c}}}_2/L^{{{\rm c}}}_1\ll\Omega^2$, the bath-induced
decoherence is slow compared to the external driving. 
In this limit, coherences can be fully established such that maximal
coherent power is injected into the system. 
Accordingly, the upper bound \eqref{PowCBound} reduces to 
$P^{{{\rm c}}}\leq -L^{{{\rm c}}}_1 F^w$, its minimum with respect to 
$\Omega$. 

The coefficients \eqref{GreenKuboCoeffC} vanish if and only if 
$G^c=0$, which means that the control variable $G^w$ commutes with the
unperturbed Hamiltonian $H$. 
Thus, according to \eqref{PowCBound}, any non-classical driving will
inevitably reduce the net output $P=P^{{{\rm d}}}+P^{{{\rm c}}}$ of 
the engine.
In fact, $P$ is subject to the upper bound 
\begin{equation}\label{PowTBound}
P\leq \frac{L_1^q F^q}{4(1+\psi_\Omega)},
\;\;\; \text{where}\;\;\;
\psi_\Omega\equiv
\frac{(L^{{{\rm c}}}_1/L^{{{\rm d}}}_1)\Omega^2}
{\Omega^2 + L^{{{\rm c}}}_2/L^{{{\rm c}}}_1}\geq 0
\end{equation}
provides a measure for the relative strength of coherent and classical
driving and 
\begin{equation}\label{TempProfSqAmpl}
F^q\equiv\frac{1}{\T}\tint (f^q_t - \bar{f}^q)^2
\quad\text{with}\quad
\bar{f}^q\equiv \frac{1}{\T}\tint f^q_t 
\end{equation}
corresponds to the mean square magnitude of the local temperature
variation induced by the heat source. 
This bound is proven in \cite{SM}.
As the bound \eqref{PowCBound}, it involves a set of 
protocol-independent parameters $L^a_j$, which are reminiscent of 
linear transport coefficients. 
For $a={{{\rm d}}}$ and $a=q$, these quantities are defined
analogously to \eqref{GreenKuboCoeffC} with $G^{{{\rm c}}}$ replaced
by $G^{{{\rm d}}}$ and $G^q$, respectively. 

In the special case of purely coherent driving, $G^{{{\rm d}}}=0$, 
the coefficient $L^{{{\rm d}}}_1$ vanishes. 
The coherence parameter $\psi_\Omega$ then diverges and
\eqref{PowTBound} reduces to $P\leq 0$.
Consequently in line with our analysis above, no cyclic engine
relying only on coherent work extraction can properly operate in the 
linear-response regime. 
For $G^{{{\rm c}}}=0$, i.e., quasi-classical driving, 
$\psi_\Omega$ vanishes and the constraint \eqref{PowTBound} assumes
its weakest form 
\begin{equation}\label{PowTClassBound}
P\leq L^q_1 F^q/4.
\end{equation}
This bound can be saturated if and only if
\begin{equation}\label{PowMaxReq}
G^w = -\mu H/T \quad\text{and}\quad
\D^\dagger H = -\lambda (H- \Sp{H}) 
\end{equation}
for some real scalars $\mu$ and $\lambda>0$, see \cite{SM}. 
Thus, the control field $f^w_t$ has to couple directly to the free 
Hamiltonian $H$, and the energy correlation function must decay
exponentially with rate $\lambda$, i.e., 
\begin{equation}\label{PowMaxDecECorr}
\bigl\llangle \hat{H}_t,\hat{H}_0\bigr\rrangle 
=e^{-\lambda t}\bigl\llangle \hat{H}_0,\hat{H}_0\bigr\rrangle.
\end{equation}
If these two requirements are fulfilled, as we show in \cite{SM}, the
protocol for optimal power extraction is determined by the condition
\begin{equation}\label{OptProt}
2\dot{f}^w_t =\lambda (f^q_t-\bar{f}^q)/\mu- \dot{f}^q_t/\mu,
\end{equation}
which leads to $P=L^q_1 F_q/4$ for any temperature profile $f^q_t$ and 
sufficiently short operation cycles 
\footnote{For long cycles, the reduced temperature profile 
$f^q_t-\bar{f}^q$ would oscillate slowly and thus be either positive
or negative over substantial time ranges. 
Consequently, integrating \eqref{OptProt} would yield a driving
protocol $f^w_t$ with large amplitude, which violates the 
linear-response condition underlying the derivations leading to 
\eqref{PowTClassBound} and \eqref{OptProt}.}.
Furthermore, using relation \eqref{PowMaxDecECorr}, the upper bound
\eqref{PowTClassBound} can be expressed in a physically transparent
way.
Specifically, we obtain
\begin{equation}\label{MaxPow}
\frac{L_1^q F^q}{4} = \lambda
\frac{\langle H^2\rangle- \Sp{H}^2}{4\kb T^3} F^q
\end{equation}
by evaluating \eqref{GreenKuboCoeffC}.
Hence, the strength and the decay rate of the energy fluctuations
in equilibrium essentially determine the maximum power output of 
a cyclic $N$-level engine in the linear-response regime. 
A similar result was obtained only recently for classical machines
obeying Fokker-Planck type dynamics \cite{Bauer2016a,Brandner2015f}. 

\begin{figure}
\includegraphics[scale=0.32]{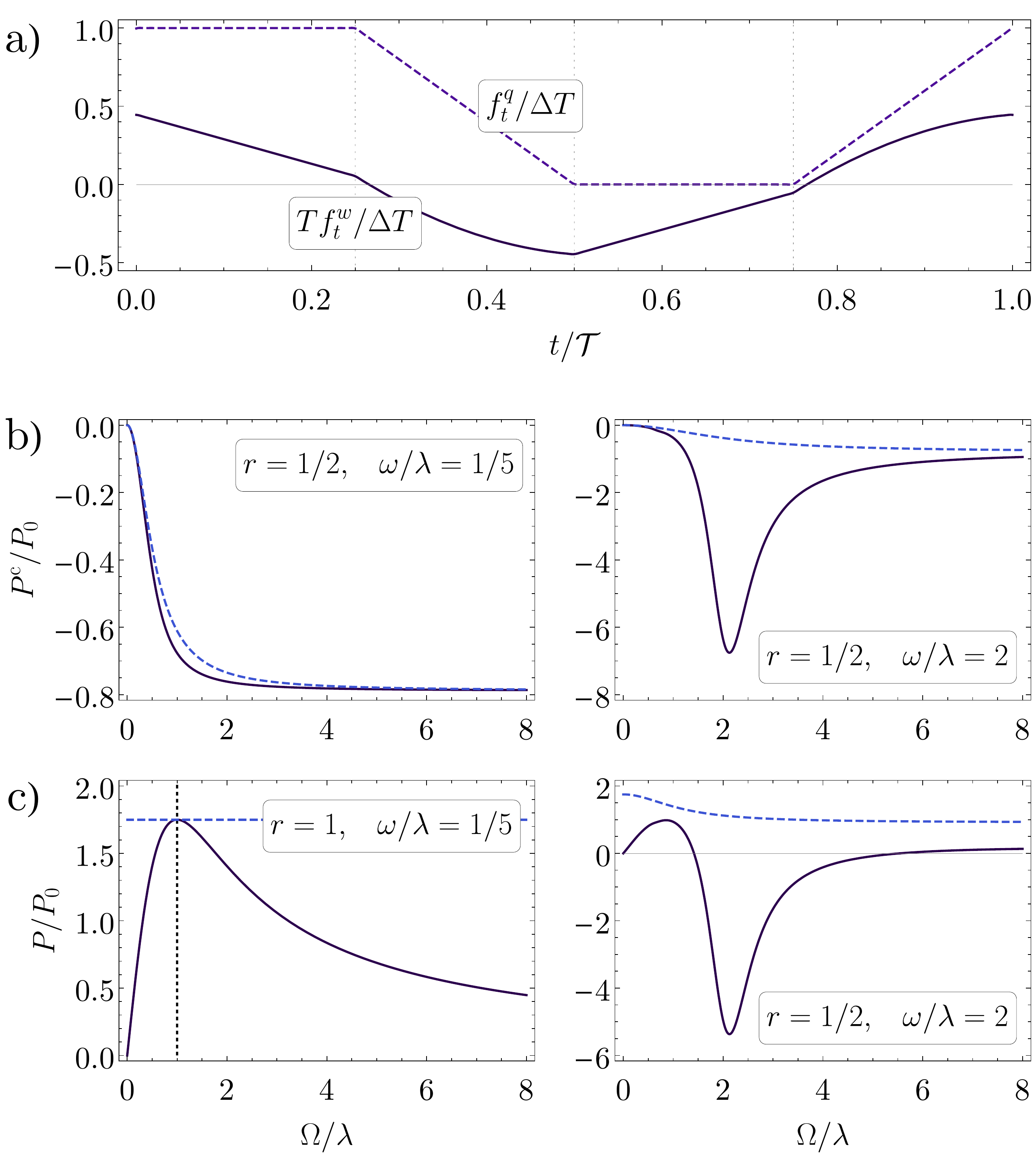}
\flushleft
\caption{Results for the single-qubit engine.
a) The temperature profile $f^q_t$ (dashed line) consists 
of two isothermal steps corresponding to the net temperatures 
$T+\Delta T$ and $T$, which are connected by linear slopes.
The work protocol $f^w_t$ (solid line) is determined by the condition 
\eqref{WorkProtTLS}. 
b) Coherent power (solid line) in units of 
$P_0\equiv (\hbar\omega\lambda/2)(\Delta T/T)^2 10^{-2}$ as a function
of the rescaled cycle frequency $\Omega/\lambda$. 
The bound \eqref{PowBoundTLS} is shown for comparison (dashed line). 
c) Plots of the total power (solid line) and its upper bound 
\eqref{PowBoundTLS} (dashed line). 
For all parts of this figure, we have set $\kappa\equiv 1$.
Symbols are explained in the main text. 
\label{FigProtPow}}
\end{figure}

We will now explore the quality of our general bounds under practical
conditions. 
To this end, we consider a two-level engine with time-dependent 
Hamiltonian
\begin{equation}\label{TLSHamiltonian}
H_t = \frac{\hbar\omega}{2}\sigma_z +\frac{\hbar\omega f^w_t}{2}
\left(r\sigma_z + (1-r)\sigma_x\right).
\end{equation}
Here, $\sigma_{x,y,z}$ are the usual Pauli matrices and the
dimensionless parameter $0\leq r \leq 1$ determines the relative
weight of the classical and the coherent parts, $G^{{{\rm d}}}=
r(\hbar\omega/2)\sigma_z$ and $G^{{{\rm c}}}= (1-r)(\hbar\omega/2)
\sigma_x$, of the control variable $G^w$. 
The corresponding equilibrium dissipator \eqref{Dissipator} involves
two Lindblad operators, $V_\pm=(\sigma_x\pm i\sigma_y)/2$, acting at
the rates $\gamma_\pm\equiv\gamma e^{\mp\kappa}$, respectively, where
$\kappa\equiv\hbar\omega\beta/2$. 
This setup lies within the range of forthcoming experiments using 
a superconducting qubit to realize the system  and ultra fast electron
thermometers for calorimetric work measurements
\cite{Gasparinetti2015,Viisanen2015,Pekola2013}.
Its coherent and total power are subject to the bounds 
\begin{align}
&P^{{{\rm c}}}\leq -\frac{\hbar\omega\lambda}{2}r^2 g\psi_\Omega F^w
\quad\text{and}\quad
P\leq \frac{\hbar\omega\lambda}{8}
\frac{g}{1+\psi_\Omega}\frac{F^q}{T^2}
\nonumber\\
&\text{with}\quad
\psi_\Omega = \frac{(1-r)^2}{r^2}\frac{\sinh 2\kappa}{4\kappa}
\frac{\Omega^2}{\Omega^2 + \omega^2 + \lambda^2/4},
\label{PowBoundTLS}
\end{align}
$g\equiv\kappa/\cosh^2\kappa$ and $\lambda\equiv 2\gamma\cosh
\kappa$, which follow from \eqref{PowCBound} and \eqref{PowTBound}
upon evaluation of the coefficients \eqref{GreenKuboCoeffC}, see
\cite{SM}.

To assess the quality of these constraints, we choose a temperature
profile $f^q_t$ that mimics the Stirling cycle illustrated in 
Fig.~\ref{FigEnConv} and a work protocol satisfying
\begin{equation}\label{WorkProtTLS}
2\dot{f}^w_t= -\Omega (f^q_t - \bar{f}^q)/T + \dot{f}^q_t/T,
\end{equation}
both shown in Fig.~\ref{FigProtPow}a. 
This choice renders the amplitude and shape of $f^w_t$ independent of
the cycle frequency $\Omega$.  
In Fig.~\ref{FigProtPow}b, the resulting coherent power is plotted 
as a function of $\Omega/\lambda$ for $r=1/2$. 
If the level splitting $\omega$ is significantly smaller than the
dissipation rate $\lambda$, it decays monotonically while closely
following its upper bound \eqref{PowBoundTLS}.
With increasing $\omega$, a resonant dip emerges close to 
$\Omega=\omega$. 
This feature is not reproduced by our bound, which is, however, still
saturated in the limits $\Omega/\lambda\rightarrow 0$ and
$\Omega/\lambda\rightarrow\infty$. 
For $r=1$, the coherent power vanishes and the two conditions
\eqref{PowMaxReq} are fulfilled with $\mu =-T$. 
The total power $P$ plotted in Fig.~\ref{FigProtPow}c then reaches its
upper bound \eqref{PowBoundTLS} at $\Omega=\lambda$, i.e., when
the work protocol \eqref{WorkProtTLS} satisfies the  maximum-power
condition \eqref{OptProt}. 
As $r$ varies from $1$ to $0$, the total power decreases more and more
due to coherence-induced losses and the bound \eqref{PowBoundTLS} lies
well above the actual value of $P$ for any cycle frequency.
This result underlines our general conclusion that coherence has a 
purely detrimental effect on power in the linear-response regime. 

For a perspective beyond linear response, we stress that our key
expressions \eqref{PowClass} and \eqref{PowQuant} are valid for
arbitrarily strong driving and any thermodynamically consistent 
time-evolution of the working fluid. 
The coherent power \eqref{PowQuant} thus constitutes a universal
indicator for the impact of quantum effects on thermal power 
generation. 
It can therefore be used as a unifying performance benchmark 
across various different types of cyclic quantum machines. 
In particular, it would be applicable to rapidly driven 
\cite{Cuetara2015,Gelbwaser-Klimovsky2015b,Alicki2015,
Gelbwaser-Klimovsky2014,Gelbwaser-Klimovsky2013,Kolar2012} and 
strongly coupled \cite{Esposito2015,Uzdin2016, Strasberg2016a,
Carrega2016a,Newman2016} engines, which are currently subject to 
active investigations. 
Furthermore, the general framework introduced in this article could 
lead to a new perspective on a phenomenon earlier interpreted as a
quantum analogue of classical friction, which was observed in models 
describing the working fluid as an interacting spin system 
\cite{Feldmann2012,Feldmann2006,Feldmann2004,Feldmann2003,
Kosloff2002}. 

As one of the earliest quantum heat engines, the three-level maser 
relies solely on non-classical work extraction \cite{Geva1994,
Scovil1959}. 
This example, which does not admit a linear-response description
\cite{Brandner2016}, shows that the coherent power can indeed become
positive if the driving is strong.
Coupled to two reservoirs with time-independent temperature, the 
three-level maser works in a steady state with respect to a rotating
basis of its Hilbert space. 
This operation principle is similar to the one used by thermoelectric
nano devices, where a spatial temperature gradient drives an electric
current \cite{Humphrey2005a}. 
Extending the concept of coherent power to this second class of 
quantum engines, which has recently attracted remarkable interest
\cite{Brandner2013,Brandner2013a,Venturelli2013,Whitney2014,
Matthews2014,Bergenfeldt2014,Hofer2015,Sanchez2015a,Brandner2015,
Marchegiani2016a,Samuelsson2016,Zheng2016} represents a challenge
promising to reveal rich and interesting physics.
Eventually, our approach could lead to a comprehensive understanding
of the role of quantum effects for one of the most fundamental
thermodynamic operations: the conversion of heat into power. 

\begin{acknowledgments}
\textbf{Acknowledgments:}
KB acknowledges financial support from the Academy of Finland
(Contract No. 296073) and is affiliated with the Centre of Quantum 
Engineering.
KB thanks J.P. Pekola, M. Campisi and R. Fazio for insightful
discussions. 
\end{acknowledgments}

\end{document}



\title{Supplemental Material for\\
Universal Coherence-Induced Power Losses of Quantum Heat 
Engines\\ in Linear Response}





\maketitle

\vbadness=10000
\hbadness=10000

\newcommand{\matt}[4]{\left(\!\begin{array}{ll}
#1 & #2\\ #3 & #4 \end{array}\!\right)}

\newcommand{\Sp}[1]{\left\langle #1 \right\rangle}
\newcommand{\Spb}[1]{\bigl\langle #1 \bigr\rangle}
\newcommand{\Spn}[1]{\langle #1 \rangle}
\newcommand{\SpB}[1]{\biggl\langle #1 \biggr\rangle}
\newcommand{\abs}[1]{\left| #1\right|}
\newcommand{\tr}[1]{{{{\rm tr}}}\!\left\{ #1 \right\}}
\newcommand{\ket}[1]{\left|#1\right\rangle}
\newcommand{\bra}[1]{\left\langle #1\right|}
\newcommand{\shortsum}[1]{\sum\mathop{}_{\mkern-5mu #1}}
\newcommand{\braket}[2]{\left\langle\left.#1\right| #2\right\rangle}

\newcommand{\x}{{{{\rm x}}}}
\renewcommand{\r}{\varrho}
\renewcommand{\j}{\mathbf{j}}
\renewcommand{\k}{\mathbf{k}}
\newcommand{\ddg}{\ddagger}

\newcommand{\kb}{k_{{{\rm B}}}}
\newcommand{\T}{\mathcal{T}}
\renewcommand{\tint}{\int_0^\T\!\!\! dt \;}
\newcommand{\tauint}{\int_0^\infty\!\!\! d\tau \;}
\newcommand{\taupint}{\int_0^\infty\!\!\! d\tau'\;}
\newcommand{\ttauint}{\int_0^\T \!\!\! dt \!
                      \int_0^\infty\!\!\! d\tau\;}

\newcommand{\K}{\mathsf{K}}
\newcommand{\s}{\sigma}
\newcommand{\vt}{\vartheta}
\newcommand{\D}{\mathsf{D}}
\renewcommand{\H}{\mathsf{H}}
\renewcommand{\L}{\mathsf{L}}
\newcommand{\Gc}{\delta G^{{{\rm c}}}}
\newcommand{\Gs}{\delta G^{{{\rm d}}}}
\newcommand{\Gq}{\delta G^q}
\newcommand{\Gw}{\delta G^w}

\makeatletter
\newcommand{\manuallabel}[2]{\def\@currentlabel{#2}\label{#1}}
\makeatother
\manuallabel{PowExpl}{11}
\manuallabel{PowCBound}{16}
\manuallabel{DB}{9}
\manuallabel{PowTBound}{19}
\manuallabel{PowTClassBound}{21}
\manuallabel{TLSHamiltonian}{26}
\manuallabel{GreenKuboCoeffC}{18}
\manuallabel{PowBoundTLS}{27}
\manuallabel{FigProtPow}{2}
\manuallabel{WorkProtTLS}{28}
\manuallabel{WorkProtSqAmpl}{17}
\manuallabel{TempProfSqAmpl}{20}

\section{Bound on Coherent Power}
We prove that the coherent power $P^{{{\rm c}}}$ as defined in 
Eq.~\ref{PowExpl} obeys the bound stated in Eq.~\ref{PowCBound}. 
To this end, it is instructive to introduce the scalar product 
\cite{Kubo1998}
\begin{equation}\label{SMInProd}
\Sp{X,Y}\equiv\int_0^\beta \!\!\! d\lambda \;
\tr{X^\dagger e^{-\lambda H} Y e^{(\lambda-\beta)H}}/
\tr{e^{-\beta H}}
\end{equation}
for any $X$ and $Y$ drawn from the space of system operators 
$\mathcal{L}$.
Furthermore, we define the super-operators 
\begin{equation}\label{SMSuperOp}
\H X \equiv [H,X]/\hbar  
\quad\text{and}\quad
\L   \equiv i\H + \D^\dagger,
\end{equation}
which have two important properties following from the detailed 
balance relation, Eq.~\ref{DB}. 
First, both, $\H$ and $\D$ are self-adjoint with respect to the inner
product \eqref{SMInProd}.
It follows that $\L^\ddagger=-i\H +\D^\dagger$, where the double
dagger indicates the super-operator adjoint with respect to
\eqref{SMInProd}. 
Second, $\H$ and $\D^\dagger$ commute. 
Therefore, $\L$ is normal, i.e., $\L\L^\ddagger = \L^\ddagger\L$
\cite{Brandner2016}. 

Using the definitions \eqref{SMInProd} and \eqref{SMSuperOp}, 
Eq.~\ref{PowExpl} can be rewritten as 
\begin{equation}\label{SMQuantPow}
P^{{{\rm c}}}= -\frac{1}{\T}\ttauint 
\dot{f}^w_t f^w_{t-\tau}\Sp{\L e^{\L\tau}\Gc,\Gc},
\end{equation}
where 
\begin{equation}\label{SMDelta}
\delta X \equiv X- \Sp{1,X}/\beta
\end{equation}
for any $X\in\mathcal{L}$. 
Due to its convolution-type structure, this expression is conveniently
analyzed in Fourier space. 
Specifically, inserting the series 
\begin{equation}
f^w_t\equiv \sum_{n\in\mathbb{Z}} c_n^w e^{in\Omega t} 
\qquad \left(\Omega\equiv 2\pi/\T\right)
\end{equation}
into \eqref{SMQuantPow} yields the mode expansion  
\begin{align}
P^{{{\rm c}}} &=
\sum_{n\in\mathbb{Z}} P^{{{\rm c}}}_n c_n^{w}c_n^{w\ast}
=\sum_{n>0}\left(P^{{{\rm c}}}_n + P^{{{\rm c}}\ast}_n\right)
c_n^{w}c_n^{w\ast}
\quad\text{with}\nonumber\\
P^{{{\rm c}}}_n & \equiv 
\Sp{\frac{in\Omega\L}{in\Omega-\L}\Gc,\Gc},
\label{SMQuantPowF}
\end{align}
$c^w_n=c^{w\ast}_{-n}\in\mathbb{C}$ and $c^w_0=0$, since the period
average of $f^w_t$ vanishes by assumption. 
The second expression is thereby obtained by formally carrying out the
improper integral in \eqref{SMQuantPow}. 
This operation is well-defined, since the super-operator $\D^\dagger$
is negative semidefinite, i.e., the eigenvalues of $\L$ have 
non-positive real part \cite{Spohn1977}.

The real part of the coefficient $P^{{{\rm c}}}_n$ defined in
\eqref{SMQuantPowF} can be bounded from above as follows. 
First, for $\mu\in\mathbb{R}$ and $n\neq 0$, we define the
quadratic form
\begin{align}
Q^{{{\rm c}}}_n(\mu)&\equiv 
\Sp{\mathsf{T}^{{{\rm c}}+}_n(\mu)\Gc ,(\L+\L^\ddagger)
\mathsf{T}^{{{\rm c}}+}_n(\mu)\Gc}\nonumber\\
&+\Sp{\mathsf{T}^{{{\rm c}}-}_n(\mu)\Gc,(\L+\L^\ddagger)
\mathsf{T}^{{{\rm c}}-}_n(\mu)\Gc}\leq 0
\quad\text{with}
\nonumber\\ 
\mathsf{T}^{{{\rm c}}\pm}_n(\mu) &\equiv
\mp\mu\frac{\L^\ddagger\mp in\Omega}{in\Omega}\pm
\frac{in\Omega}{\L\pm in\Omega}.
\label{SMQcDef}
\end{align}
Note that $Q^{{{\rm c}}}_n(\mu)\leq 0$, since the super-operator 
$\L+\L^\ddagger = 2\D^\dagger$ is negative semidefinite
\cite{Spohn1977}. 
Second, we observe that 
\begin{align}
&\Spn{Y^\dagger,Z^\dagger}=\Sp{Y,Z}^\ast = \Sp{Z,Y}
\label{SMSymProp1}\quad\text{and}\\
&(\L Y)^\dagger = \L Y^\dagger,\quad
(\L^\ddagger Y)^\dagger = \L^\ddagger Y^\dagger
\label{SMSymProp2}
\end{align}
for arbitrary operators $X,Y\in\mathcal{L}$. 
Using these relations and the fact that the operator $\Gc$ is
Hermitian, the quadratic form \eqref{SMQcDef} can be expanded as 
\begin{align}
Q_n^{{{\rm c}}}(\mu)&= 
4\mu^2 \Sp{\L\Gc,\L^2\Gc}/(n\Omega)^2+4\mu^2\Sp{\Gc,\L\Gc}
\nonumber\\
&+8\mu \Sp{\Gc,\L\Gc} 
+2\left( P^{{{\rm c}}}_n +P^{{{\rm c}}\ast}_n \right)\leq 0. 
\label{SMQcExpand}
\end{align}
Maximizing this expression with respect to $\mu$ yields
\begin{align}\label{SMQcFin}
P^{{{\rm c}}}_n + P^{{{\rm c}}\ast}_n
&\leq\frac{2\Sp{\L\Gc,\Gc}^2}{\Sp{\L\Gc,\Gc}
+\Sp{\L\Gc,\L^2\Gc}/(n\Omega)^2}\nonumber\\
&\leq\frac{2\Sp{\L\Gc,\Gc}^2}{\Sp{\L\Gc,\Gc}
+\Sp{\L\Gc,\L^2\Gc}/\Omega^2}
\leq 0. 
\end{align}

We now observe that 
\begin{align}
-\Sp{\L\Gc,\Gc} &= 
\int_0^\infty\!\!\! dt\; \Sp{e^{\L t} \L\Gc,\L\Gc}\nonumber\\
&= \int_0^\infty\!\!\! dt\; \bigl\llangle 
\hat{G}^{{{\rm c}}(1)}_t, \hat{G}^{{{\rm c}}(1)}_0\bigr\rrangle
=L^{{{\rm c}}}_1 \geq 0. 
\label{SMLc1Def}
\end{align}
and
\begin{align}
-\Sp{\L\Gc,\L^2\Gc} &=
\int_0^\infty\!\!\! dt\; \Sp{e^{\L t}\L^2\Gc,\L^2\Gc}\nonumber\\
&=\int_0^\infty\!\!\! dt\; \bigl\llangle\hat{G}^{{{\rm c}}(2)}_t,
\hat{G}^{{{\rm c}}(2)}_0\bigr\rrangle = L^{{{\rm c}}}_2\geq 0.
\label{SMLc2Def}
\end{align}
Hence, the bound \eqref{SMQcFin} can be cast into the compact form
\begin{equation}\label{SMPcBup}
P^{{{\rm c}}}_n+P^{{{\rm c}}\ast}_n \leq 
-\frac{2L^{{{\rm c}}}_1\Omega^2}{
\Omega^2+L_2^{{{\rm c}}}/L^{{{\rm c}}}_1}. 
\end{equation}
Using this result to bound the mode expansion
\eqref{SMQuantPowF} yields 
\begin{align}
P^{{{\rm c}}} &\leq -\frac{2L^{{{\rm c}}}_1\Omega^2}{
\Omega^2+L_2^{{{\rm c}}}/L^{{{\rm c}}}_1}
\sum_{n>0}c_n^w c_n^{w\ast}\nonumber\\
&=-\frac{L^{{{\rm c}}}_1\Omega^2}{
\Omega^2+L_2^{{{\rm c}}}/L^{{{\rm c}}}_1}
\frac{1}{\T}\tint (f^w_t)^2\leq 0,
\label{SMQuantPowBnd}
\end{align}
i.e., the upper bound stated in Eq.~\ref{PowCBound}. 

We note that, first, for $\Omega\rightarrow 0$, the
coefficients $P_n^{{{\rm c}}}$ defined in \eqref{SMQuantPowF} vanish
such that $P^{{{\rm c}}}\rightarrow 0$.  
Second, \eqref{SMQcDef} and \eqref{SMQcExpand} imply 
\begin{equation}
\lim_{\Omega\rightarrow\infty} Q^{{{\rm c}}}_n(\mu=0) 
= -4L^{{{\rm c}}}_1 = 
2\left(P^{{{\rm c}}}_n + P^{{{\rm c}}\ast}_n\right).
\end{equation}
Recalling \eqref{SMQuantPowF}, we thus obtain
\begin{equation}
\lim_{\Omega\rightarrow\infty} P^{{{\rm c}}} 
= -2 L^{{{\rm c}}}_1\sum_{n>0} c_n^w c_n^{w\ast} 
= -\frac{L^{{{\rm c}}}_1}{\T}\tint (f^w_t)^2.
\end{equation}
Consequently, the bound \eqref{SMQuantPowBnd} is saturated in the 
two limits $\Omega\rightarrow 0$ and $\Omega\rightarrow\infty$. 
Finally, the coefficients $L^{{{\rm c}}}_j$ defined in
\eqref{SMLc1Def} and \eqref{SMLc2Def} vanish if and only if $\Gc=0$, 
because all eigenvalues of $\L$ have non-positive real part and the
operator $\Gc$ is orthogonal to the null space of $\L$. 
Since the set of Lindblad-operators $\{V_\sigma\}$ is self-adjoint and
irreducible, this space contains only scalar multiples of the identity
operator \cite{Spohn1977}.

\section{Bound on Total Power}
The upper bound on the total power output stated in 
Eq.~\ref{PowTBound} can be established in two major steps. 
First, we observe that using Eq.~\ref{PowExpl} and the notation
introduced in the previous section, $P=P^{{{\rm c}}}+P^{{{\rm d}}}$
can be written as 
\begin{multline}
P= -\frac{1}{\T}\ttauint\Bigl(
\dot{f}^w_t f^w_{t-\tau}\Sp{\L e^{\L\tau}\Gc,\Gc}\\
+\dot{f}^w_t f^w_{t-\tau}\Sp{\L e^{\L\tau}\Gs,\Gs}
+\dot{f}^w_t f^q_{t-\tau}\Sp{\L e^{\L\tau}\Gs,\Gq}\Bigr).
\end{multline}
Upon insertion of the Fourier series expansion 
\begin{equation}\label{SMProtF}
f^a_t\equiv\sum_{n\in\mathbb{Z}}c_n^a e^{in\Omega t}
\qquad\left(\Omega\equiv 2\pi/\T\right)
\end{equation}
with $c_n^a=c_{-n}^{a\ast}\in\mathbb{C}$ and $a=w,q$, this expression 
becomes 
\begin{align}\label{SMTPowF}
P &=\sum_{n\in\mathbb{Z}}
\left(P^{{{\rm d}}}_n +P^{{{\rm c}}}_n\right)c_n^w c_n^{w\ast} 
+ P^q_n c_n^w c_n^{q\ast}\nonumber\\
&=\sum_{n>0}\left(1+\frac{P^{{{\rm c}}}_n+P^{{{\rm c}}\ast}_n}
{P^{{{\rm d}}}_n+P^{{{\rm d}}\ast}_n}\right)
\left(P^{{{\rm d}}}_n+P^{{{\rm d}}\ast}_n\right) c^w_n c^{w\ast}_n
\nonumber\\
&\hspace*{3.9cm}
+ P^q_n c_n^w c_n^{q\ast}+ P^{q\ast}_n c_n^{w\ast} c_n^q,
\end{align}
where 
\begin{align}
P^{{{\rm d}}}_n &\equiv\SpB{\frac{in\Omega\L}{in\Omega-\L}\Gs,\Gs}
=\SpB{\frac{in\Omega\D^\dagger}{in\Omega-\D^\dagger}\Gs,\Gs},
\label{SMTPowFCoeff1}\\
P^{q}_n & \equiv\SpB{\frac{in\Omega\L}{in\Omega-\L}\Gs,\Gq}
=\SpB{\frac{in\Omega\D^\dagger}{in\Omega-\D^\dagger}\Gs,\Gq}
\label{SMTPowFCoeff2}
\end{align}
and $P^{{{\rm c}}}_n$ was defined in \eqref{SMQuantPowF}. 
In \eqref{SMTPowFCoeff1} and \eqref{SMTPowFCoeff2}, we
could replace $\L=i\H+\D^\dagger$ by $\D^\dagger$ since, by 
construction, both $\Gs$ and $\Gq$ commute with $H$, i.e., 
$\H\Gs=\H\Gq=0$. 
We now observe that the real part of the coefficient 
$P_n^{{{\rm d}}}$ obeys 
\begin{equation}\label{SMPsBlow}
-2L^{{{\rm d}}}_1
\leq P^{{{\rm d}}}_n+P^{{{\rm d}}\ast}_n\leq 0,
\end{equation}
where, analogous to \eqref{SMLc1Def}, 
\begin{equation}
L_1^{{{\rm d}}}\equiv \int_0^\infty\!\!\! dt\; \bigl\llangle 
\hat{G}^{{{\rm d}}(1)}_t,\hat{G}^{{{\rm d}}(1)}_0\bigr\rrangle\geq 0.
\end{equation}
This constraint can be derived by repeating the steps \eqref{SMQcDef}
to \eqref{SMPcBup} with the quadratic from 
\begin{align}
Q^{{{\rm d}}}_n(\mu)&\equiv
\Sp{\mathsf{T}^{{{\rm d}}}_n(\mu)
\Gs,\D^\dagger\mathsf{T}^{{{\rm d}}}_n(\mu)\Gs}\leq 0,
\quad\text{where}\nonumber\\
\mathsf{T}^{{{\rm d}}}_n(\mu) &\equiv\mu
+ \frac{in\Omega}{\D^\dagger + in\Omega}
\quad\text{and}\quad
n\neq 0, \; \mu\in\mathbb{R}.
\label{SMQsDef}
\end{align}
With \eqref{SMPcBup}, \eqref{SMPsBlow} implies 
\begin{equation}
1 +\frac{P^{{{\rm c}}}_n+P^{{{\rm c}}\ast}_n}
{P^{{{\rm d}}}_n+P^{{{\rm d}}\ast}_n}\geq 1+
\frac{(L^{{{\rm c}}}_1/L^{{{\rm d}}}_1)\Omega^2}
{\Omega^2 + L^{{{\rm c}}}_2/L^{{{\rm c}}}_1}
\equiv\phi_\Omega\geq 1 
\end{equation}
and thus, recalling \eqref{SMTPowF}, 
\begin{equation}\label{SMTPowFRed}
P\leq\sum_{n>0}\phi_\Omega
\left(P^{{{\rm d}}}_n+P^{{{\rm d}}\ast}_n\right) c^w_n c^{w\ast}_n
+ P^q_n c_n^w c_n^{q\ast}+ P^{q\ast}_n c_n^{w\ast} c_n^q. 
\end{equation}

For the second step of our analysis, we note that the inequality
\eqref{SMTPowFRed} can then be rewritten as 
\begin{align}
P&\leq \sum_{n>0}\phi_\Omega
\Sp{\frac{in\Omega\D^\dagger}{in\Omega-\D^\dagger}\Gs
   +\frac{in\Omega\D^\dagger}{in\Omega+\D^\dagger}\Gs,\Gs}
c_n^wc_n^{w\ast}\nonumber\\
&\hspace*{1.5cm}
+\Sp{\frac{in\Omega\D^\dagger}{in\Omega-\D^\dagger}\Gs,\Gq}
c_n^w c_n^{q\ast}\nonumber\\
&\hspace*{1.5cm}
+\Sp{\frac{in\Omega\D^\dagger}{in\Omega+\D^\dagger}\Gs,\Gq}
c_n^{w\ast} c_n^q\\[9pt]
&=\sum_{n>0}
2\phi_\Omega\Sp{K_n,\D^\dagger K_n} 
-\frac{\Sp{\Gq,\D^\dagger\Gq}}{2\phi_\Omega}c_n^qc_n^{q\ast}
\label{SMTPowFBound}
\end{align}
with
\begin{equation}
K_n\equiv c_n^w\frac{in\Omega}{\D^\dagger+in\Omega}\Gs
+\frac{c_n^q}{2\phi_\Omega}\Gq. 
\end{equation}
Here, we have exploited that $\Gs$ and $\Gq$ are Hermitian operators,
the relations \eqref{SMSymProp1} and \eqref{SMSymProp2} and the fact
that the super-operator $\D^\dagger$ is self-adjoint. 
Moreover, since, $\D^\dagger$ is negative semidefinite, the first 
term under the sum in \eqref{SMTPowFBound} is non-positive, while the
second one is non-negative. 
Thus, we have 
\begin{align}
P&\leq -\frac{\Sp{\Gq,\D^\dagger\Gq}}{2\phi_\Omega}
\sum_{n>0}c_n^qc_n^{q\ast}\nonumber\\
&= -\frac{\Sp{\Gq,\L\Gq}}{4\phi_\Omega}\frac{1}{\T}
\tint\left(f^q_t-\bar{f}^q\right)^2.
\end{align}
Finally, making the identification
\begin{align}
-\Sp{\Gq,\L\Gq}=\int_0^\infty\!\!\! dt\;\bigl\llangle
\hat{G}^{q(1)}_t,\hat{G}^{q(1)}_0\bigr\rrangle\equiv L^q_1
\end{align}
and defining $\psi_\Omega\equiv \phi_\Omega -1$ completes our proof
of Eq.~\ref{PowTBound}. 

\section{Optimal Classical Driving}
We investigate the conditions that allow saturation of the 
quasi-classical bound on the total power output stated in 
Eq.~\ref{PowTClassBound}. 
To this end, using the previously introduced notation, the deviation 
of $P$ from its upper limit $L_1^q F_q/4$ can be written in the form 
\begin{equation}\label{SMPowTClassSat1}
P-L_1^q F_q/4 = 2\sum_{n>0}\Sp{K'_n,\D^\dagger K'_n}
\end{equation}
with
\begin{equation}\label{SMPowTClassSat2}
K'_n = c_n^w\frac{in\Omega}{\D^\dagger + in\Omega}\Gw
+ \frac{c_n^q}{2}\Gq.
\end{equation}
This expression can be obtained by repeating the derivation of 
\eqref{SMTPowFBound} and invoking the additional condition $\Gc=0$,
which implies $\Gs=\Gw$. 

From \eqref{SMPowTClassSat1} and \eqref{SMPowTClassSat2}, it follows
that $P=L^q_1F_q/4$ if and only if, each $K'_n$ lies 
in the null space of the super-operator $\D^\dagger$, i.e., if 
$K'_n$ is a scalar multiple of the identity operator for any 
integer $n>0$. 
This condition requires 
\begin{equation}\label{SMPowTClassSat3}
0= 2in\Omega c_n^w\Gw + in\Omega c_n^q\Gq + c_n^q\D^\dagger\Gq. 
\end{equation}
We now observe that, since $\Gw$ and $\Gq$ must be Hermitian 
operators, \eqref{SMPowTClassSat3} has a non-trivial solution only if
\begin{equation}\label{SMPowTClassSat4}
\Gw=\mu\Gq \quad\text{and}\quad \D^\dagger\Gq=-\lambda\Gq
\end{equation} 
for some real $\mu$ and $\lambda>0$, i.e., if the variable 
$G^w$ is proportional to the unperturbed Hamiltonian $H$ and $\Gq$
is an eigenvector of $\D^\dagger$. 
This connection can be established by splitting the Fourier 
coefficients $c_n^w$ and $c_n^q$ in real and imaginary parts and 
therewith separating \eqref{SMPowTClassSat3} into two linear equations
with real coefficients. 
Recalling the definition \eqref{SMDelta} and that $G_q = -H/T$,
\eqref{SMPowTClassSat4} can be written as 
\begin{equation}
G^w = -\mu H/T\quad\text{and}\quad
\D^\dagger H = -\lambda(H-\Sp{H}).
\end{equation}

Provided the two conditions \eqref{SMPowTClassSat4} are met, 
the solution of \eqref{SMPowTClassSat3} reads 
\begin{equation}
2 in\Omega c_n^w = \lambda c_n^q/\mu -in\Omega c_n^q/\mu.
\end{equation}
Inverting the Fourier transformation \eqref{SMProtF} thus yields the
differential equation
\begin{equation}
2\dot{f}^w_t = \lambda(f^q_t -\bar{f}^q)/\mu  -\dot{f}^q_t/\mu.
\end{equation}

\section{Single-Qubit Engine}
We consider the two-level engine described by the Hamiltonian shown 
in Eq.~\ref{TLSHamiltonian}. 
For this system, the super-operator $\L$ defined in \eqref{SMSuperOp}
has the explicit form 
\begin{multline}
\L X = \frac{i\omega}{2}[\sigma_z,X] 
+ \frac{\gamma e^{\kappa}}{2}\left(V_+[X,V_-]
+ [V_+,X]V_-]\right)\\
+ \frac{\gamma e^{-\kappa}}{2}\left(V_-[X,V_+]
+ [V_-,X]V_+]\right)
\end{multline}
with $V_\pm\equiv (\sigma_x\pm i\sigma_y)/2$.
It fulfills 
\begin{align}
\L\sigma_z &= -\lambda\left(\sigma_z +\tanh\kappa\right)
            = -\lambda\left(\sigma_z - \Sp{\sigma_z}\right),
\nonumber\\ 
\L V_\pm & = \left(\pm i\omega -\lambda/2\right)V_\pm. 
\end{align}
Using these relations and the fact that $\sigma_x = V_++V_-$, it is
straightforward to evaluate the coefficients $L^a_j$ defined in
Eq.~\ref{GreenKuboCoeffC}.
Specifically, we find 
\begin{align}
L^{{{\rm c}}}_j &= \frac{\hbar\omega\lambda (1-r)^2\tanh\kappa}{4}
(\omega^2 + \lambda^2/4)^{j-1}, \nonumber\\
L^{{{\rm d}}}_j &= \frac{\hbar\omega\lambda r^2\kappa}{2\cosh^2\kappa}
\lambda^{2(j-1)}, \nonumber\\
L^{q}_j &= \frac{\hbar\omega\lambda\kappa}{2\cosh^2\kappa}
\frac{\lambda^{2(j-1)}}{T^2}. 
\end{align}
Plugging these results into the general bounds given in 
Eqs.~\ref{PowCBound} and \ref{PowTBound} yields Eq.~\ref{PowBoundTLS}. 

For the plots of Fig.~\ref{FigProtPow}, we use the piecewise 
defined temperature profile 
\begin{equation}
f^q_t = \Delta T\begin{cases} 1, & t\leq \T/4\\
                              2-4t/\T, & \T/4\leq t\leq \T/2\\
                              0, & \T/2\leq t\leq 3\T/4\\
                              4t/\T-3, & 3\T/4\leq t\leq \T
\end{cases}
\end{equation}
with Fourier coefficients 
\begin{equation}
c_0^q=\Delta T/2, \quad
c_n^q=-\Delta T\frac{(-1)^{n}(i^n+1)(i^n-1)^2}{n^2\pi^2}
\quad (n\neq 0),
\end{equation}
see \eqref{SMProtF}. 
The Fourier coefficients of the work protocol defined through the
conditions Eq.~\ref{WorkProtTLS} and $\tint f^w_t =0$ are given by
$c^w_0=0$ and 
\begin{equation}
c^w_n=\frac{1}{T}\frac{in-1}{2in}c_n^q
\quad (n\neq 0). 
\end{equation}
Evaluating the corresponding mean square amplitudes, which were 
defined in Eqs.~\ref{WorkProtSqAmpl} and  \ref{TempProfSqAmpl}, yields
\begin{align}
F^w &= 2\sum_{n>0} c_n^w c_n^{w\ast} = \frac{\Delta T^2}{T^2}
\frac{\pi^2+10}{240}\quad\text{and}\nonumber\\
F^q &= 2\sum_{n>0} c_n^q c_n^{q\ast} = \frac{\Delta T^2}{6}.
\end{align}
Furthermore, the mode expansion coefficients of the coherent and the
classical power, which were introduced in \eqref{SMQuantPowF},
\eqref{SMTPowFCoeff1} and  \eqref{SMTPowFCoeff2}, respectively, 
become
\begin{align}
P^{{{\rm c}}}_n &=\frac{\hbar\omega (1-r)^2\tanh\kappa}{4}
\nonumber\\
&\hspace*{2cm} \times \left(\frac{in\Omega (i\omega-\lambda/2)}
{in\Omega+(i\omega-\lambda/2)}- \frac{in\Omega (i\omega+\lambda/2)}
{in\Omega-(i\omega+\lambda/2)}\right),\nonumber\\
P^{{{\rm d}}}_n &= \frac{\hbar\omega r^2 \kappa}{2\cosh^2\kappa}
\frac{in\Omega\lambda}{\lambda-in\Omega},\nonumber\\
P^q_n & = -\frac{1}{T}\frac{\hbar\omega r\kappa }{2\cosh^2\kappa}
\frac{in\Omega\lambda}{\lambda-in\Omega}.
\end{align}
The plots shown in Fig.~\ref{FigProtPow} are now obtained by 
numerically approximating the improper sums in \eqref{SMQuantPowF}
and \eqref{SMTPowF}.